\documentclass[12p]{article}
\usepackage{latexsym}
\usepackage{color}
\usepackage{amsmath}
\usepackage{epsfig}
\usepackage{hyperref}
 \usepackage{dcolumn}
   \usepackage{threeparttable}
     \newcommand{\thickhline}{\noalign{\hrule height 0.8pt}}
\newcommand{\ortala}[1]{\begin{center}#1\end{center}}

\newcommand{\sandd}[1]{\left\langle #1\right\rangle}

\newcommand{\integ}[3]{{{\underset{#1 }{\overset{#2}{\displaystyle\int}}}#3}}

\newcommand{\summ}[3]{{{\underset{#1 }{\overset{#2}{\displaystyle\sum}}}#3}}

\newcommand{\re}[1]{(\ref{#1})}

\newcommand{\eq}[2]{\begin{equation}\label{#1}  #2\end{equation}}

\newcommand{\paran}[1]{\left(#1\right)}

\newcommand{\sch}[1]{Schrodinger}

\newcommand{\komb}[2]{\paran{\begin{array}{c} #1 \\ #2 \end{array}}}

\linethickness{0.6pt}

\begin{document}

\ortala{\textbf{Random field effects on the isotropic quantum Heisenberg model with Gaussian random magnetic field distribution}}

\ortala{\textbf{\"Umit Ak\i nc\i \footnote{umit.akinci@deu.edu.tr}}}

\ortala{\textit{Department of Physics, Dokuz Eyl\"ul University,
TR-35160 Izmir, Turkey}}

\section{Abstract}

Effect of Gaussian random magnetic field distribution which is centered at zero on the phase transition properties of isotropic quantum Heisenberg model has been
investigated on two (2D) and three dimensional (3D) lattices within the framework of effective field theory (EFT) for a two spin cluster (which is abbreviated as EFT-2). Beside the phase diagrams and the evolution of the magnetization versus temperature curves with the Gaussian magnetic field distribution width, critical Gaussian distribution width values, which make the critical temperature zero, have been obtained for several lattices. Moreover, it has been concluded that all critical temperatures are of the second order and  reentrant behavior does not exist in the phase  diagrams.

Keywords: \textbf{Quantum isotropic Heisenberg model; Random
magnetic field; Gaussian magnetic distribution}
\section{Introduction}\label{introduction}


Recently, there has been growing theoretical interest in the random field lattice spin models.  The model was introduced for the first time by Larkin \cite{ref1} for superconductors and later generalized by Imry and Ma \cite{ref2}.
Ising model which is the most basic lattice spin model with  quenched random field (RFIM) has been studied over three decades, since this model can be used to describe a wide variety of disordered systems.
Diluted antiferromagnets (such as $Fe_xZn_{1-x}F_2$, $Rb_2Co_xMg_{1-x}F_4$ and $Co_xZn_{1-x}F_2$)
in a homogenous magnetic field behave like ferromagnetic systems in the presence of random fields \cite{ref3,ref4}.
Structural phase transitions in random alloys, commensurate charge-density-wave systems with impurity pinning, binary fluid mixtures in random porous
media, and the melting of intercalates in layered compounds, such as $TiS_2$ \cite{ref5} are the examples of the experimentally accessible disordered systems which can be described by RFIM.
Besides, RFIM can mimic the phase transitions and interfaces in random media
\cite{ref6,ref7}, e.g pre-wetting transition on a disordered substrate can be mapped onto a
2D RFIM problem \cite{ref8}. RFIM has also been applied to describe critical
surface behavior of amorphous semi-infinite systems \cite{ref9,ref10}.

RFIM has been widely studied  in the literature with discrete \cite{ref11,ref12,ref13,ref14,ref15}, as well as continous  \cite{ref16,ref17,ref18,ref19,ref20,ref21} distributions. Random distribution of the magnetic field may produce drastic effects on the phase
diagrams and related magnetic properties of the system. It has been shown that Ising systems under
the influence of discrete symmetric distributions,
like bimodal \cite{ref11} and trimodal \cite{ref12} distributions, exhibit tricritical behavior,
while  continuous symmetric distributions like Gaussian distribution \cite{ref16} lead to only second order transitions.

Although the results of the RFIM have been reported for both discrete and continuous distributions, there has been less attention
paid on the random field effects on the Heisenberg model. Since Heisenberg model is more realistic model than the Ising model for the spin systems, it is important  to investigate this model in the presence of quenched random fields. Albuquerque and Arruda \cite{ref22} studied the effect of the bimodal random field distribution on the phase transition characteristics of the spin-1/2 isotropic classical  Heisenberg model and they found tricritical behavior within the EFT-2 formulation. Oubelkacem et al., studied the same system with another approach, namely EFT with probability distribution technique and they obtained similar results \cite{ref23}.
Albuquerque et al. \cite{ref24} treated the same system with amorphization effect, again with the EFT-2 formulation.  Recently,
Sousa et al. have studied the effect of the bimodal random field distribution on phase transition characteristics of the isotropic classical- and quantum- spin-1/2 Heisenberg model within the EFT-2 formulation and also they found a tricritical behavior \cite{ref25}. All of these works have been restricted to the spin-1/2 isotropic Heisenberg model with bimodal random field distribution and they concluded that tricritical behavior exists in this system as in Ising model with bimodal random field distribution. They utilized an EFT formalism which is characterized by differential operator technique introduced by Honmura and Kaneyoshi for Ising systems \cite{ref26}.

EFT approximation can provide results that are superior to those obtained
within the traditional mean field approximation, due to the consideration of self spin correlations which are omitted in the mean field approximation. EFT for a typical Ising system starts by constructing a finite cluster of spins which represents the system. Callen-Suzuki spin identities \cite{ref27,ref28} are the starting point of the EFT for the one spin clusters. If one expands these identities with differential operator technique, multi spin correlations appear, and in order to avoid from the mathematical difficulties, these multi spin correlations are often neglected by using a decoupling approximation \cite{ref29}. Working with larger finite clusters will give more accurate results. Callen-Suzuki identities have been generalized to two spin clusters in Ref. \cite{ref30} (EFT-2 formulation). This EFT-2 formulation has been successfully applied to a variety of systems, such as quantum spin-1/2 Heisenberg ferromagnet \cite{ref31,ref32}  and antiferromagnet \cite{ref33} systems,  classical n-vector model \cite{ref34,ref35}, and spin-1 Heisenberg ferromagnet \cite{ref36,ref37}.

The aim of this work is to investigate the effect of the Gaussian random field distribution centered at zero on the phase transition characteristics of a spin-1/2 isotropic quantum Heisenberg model. Quantum Heisenberg model can take into account the quantum fluctuations which dominates the thermal fluctuations in the low temperatures. Thus, it is expected that it gives more reasonable results than the classical one in this low temperature region. We follow the EFT-2 formulation which is derived in Ref. \cite{ref31} for this system.

The paper is organized as follows: In Sec. \ref{formulation}, we
briefly present the model and  formulation. The results and
discussions are presented in Sec. \ref{results}, and finally Sec.
\ref{conclusion} contains our conclusions.

\section{Model and Formulation}\label{formulation}

We consider a lattice which consists of $N$ identical spins (spin-$1/2$) such that each of the spins has $z$ nearest neighbors. The Hamiltonian of the system is given by
\eq{denk1}{\mathcal{H}=-J\summ{<i,j>}{}{\mathbf{s}_i.\mathbf{s}_j}-\summ{i}{}{H_is_i^z}}
where $\mathbf{s}_i$ and $s_i^z$  denote the Pauli spin operator and the $z$ component of the Pauli spin operator at a site $i$, respectively. $J$ stands for the exchange interactions between the nearest neighbor spins and $H_i$ is the longitudinal magnetic field acting on the site $i$. The first summation is carried over the nearest neighbors of the lattice, while the second one is over all the lattice sites. Magnetic fields are distributed on the lattice sites according to a Gaussian distribution
function which is given by
\eq{denk2}{P\paran{H_i}=
\frac{1}{\sqrt{2\pi \sigma^2}}\exp{\paran{-\frac{H_i^2}{2\sigma^2}}}
} where $\sigma$ is the width of the distribution. According to Eq. \re{denk2}, the average magnetic field value of the overall system is zero.  The limit $\sigma \rightarrow 0$ covers a pure system, i.e. Heisenberg model under zero magnetic field.

We use the two spin cluster approximation as an EFT formulation, namely EFT-2 formulation \cite{ref31}. In this approximation, we select two spins (namely $s_1$ and $s_2$) and treat the interactions exactly in this two spin cluster. In order to avoid some mathematical difficulties, we replace the perimeter spins of the two spin cluster by Ising spins (axial approximation) \cite{ref32}. After all, by using the differential operator technique with decoupling approximation \cite{ref29},
we get an expression for the magnetization per spin as
\eq{denk3}{
m=\sandd{\frac{1}{2}\paran{s_1^z+s_2^z}}=\sandd{\left[A_{x}+m B_{x}\right]^{z_0}
\left[A_{y}+m B_{y}\right]^{z_0}
\left[A_{xy}+m B_{xy}\right]^{z_1}} F\paran{x,y}|_{x=0,y=0}
} where $s_1$ and $s_2$ have $z_0$ distinct nearest neighbors and both of them have $z_1$ common nearest neighbors.

The coefficients are defined by
\eq{denk4}{
\begin{array}{lcl}
A_{x}=\cosh{\paran{J_z\nabla_x}}&\quad&
B_{x}=\sinh{\paran{J_z\nabla_x}}\\
A_{y}=\cosh{\paran{J_z\nabla_y}}
&\quad&
B_{y}=\sinh{\paran{J_z\nabla_y}}\\
A_{xy}=\cosh{\left[J_z\paran{\nabla_x+\nabla_y}\right]}&\quad&
B_{xy}=\sinh{\left[J_z\paran{\nabla_x+\nabla_y}\right]}\\
\end{array}
}
where $\nabla_x=\partial/\partial x$ and $\nabla_y=\partial/\partial y$ are the usual differential operators in the
differential operator technique. Differential operators act on an arbitrary function via
\eq{denk5}{\exp{\paran{a\nabla_x+b\nabla_y}}G\paran{x,y}=G\paran{x+a,y+b}}
with any constant  $a$ and $b$. The function in Eq. \re{denk3} is given by
\eq{denk6}{F\paran{x,y}=\integ{}{}{}dH_1dH_2P\paran{H_1}P\paran{H_2}f\paran{x,y,H_1,H_2}}
where
\eq{denk7}{f\paran{x,y,H_1,H_2}=\frac{\sinh{\paran{\beta X_0}}}{\cosh{\paran{\beta X_0}}+\exp{\paran{-2\beta J}}\cosh{\paran{\beta Y_0}}}} and
\eq{denk8}{
X_0=(x+y+H_1+H_2), \quad Y_0=\left[\paran{2J}^2+(x-y+H_1-H_2)^2\right]^{1/2},
}
with $\beta=1/(k_B T)$, $k_B$ is the Boltzmann
constant and $T$ is the temperature. With the help of the Binomial expansion, Eq. \re{denk3} can be written as
\eq{denk9}{
m=\summ{p=0}{z_0}{}\summ{q=0}{z_0}{}\summ{r=0}{z_1}{}C^\prime_{pqr}m^{p+q+r}
} where the coefficients are
\eq{denk10}{
C^\prime_{pqr}=\komb{z_0}{p}\komb{z_0}{q}\komb{z_1}{r}A_x^{z_0-p}A_y^{z_0-q}A_{xy}^{z_1-r}B_x^{p}B_y^{q}B_{xy}^{r}F\paran{x,y}|_{x=0,y=0}
} and these coefficients can be calculated by using the definitions given in Eqs. \re{denk4} and \re{denk5}. Let us write Eq. \re{denk9} in more familiar form as
\eq{denk11}{
m=\summ{k=0}{z}{}C_{k}m^{k}
} and
\eq{denk12}{
C_{k}=\summ{p=0}{z_0}{}\summ{q=0}{z_0}{}\summ{r=0}{z_1}{}\delta_{p+q+r,k}C^\prime_{pqr}
} where $\delta_{i,j}$ is the Kronecker delta. It can be shown from the symmetry properties of the function defined in Eq. \re{denk6} and operators defined by Eq. \re{denk4} that for even $k$, the coefficient $C_k$ is equal to zero. This property is derived in Sec. \ref{sec_app_a}.

For a given set of Hamiltonian parameters ($J$), temperature ($k_BT/J$) and field distribution parameter ($\sigma$), we can determine the coefficients  from Eq. \re{denk12} and we can obtain
a non linear equation from Eq. \re{denk11}. By solving this equation, we can get the magnetization ($m$) for a given set of parameters and temperature. Since the magnetization is close to zero in the vicinity of the critical point, we can obtain a linear equation by linearizing the equation given in  Eq. \re{denk11} which allows us to determine the critical temperature. Since we have not calculated the free energy in this approximation, we can locate only second order transitions from the condition
\eq{denk13}{
C_1=1, \quad C_3<0.
}The tricritical point at which the second and first order transition lines meet can be determined from the condition
\eq{denk14}{
C_1=1, \quad C_3=0.
}

\section{Results and Discussion}\label{results}

The effect of the zero centered Gaussian magnetic field distribution on the isotropic quantum Heisenberg model problem has one parameter as a measure of randomness, namely the width of the distribution $\sigma$. As $\sigma$ increases then randomly distributed magnetic fields with greater absolute strengths start to act on the lattice sites.

\begin{figure}[h]\begin{center}
\epsfig{file=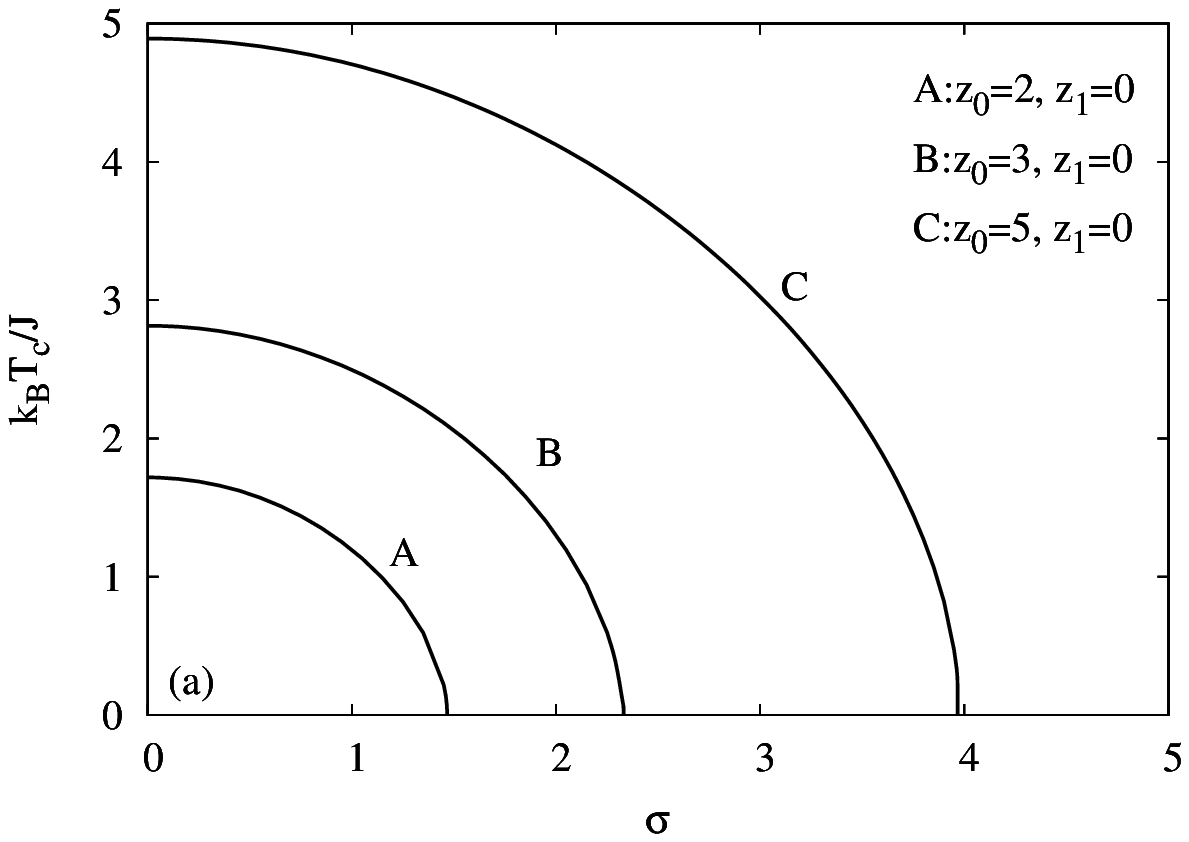, width=6.5cm}
\epsfig{file=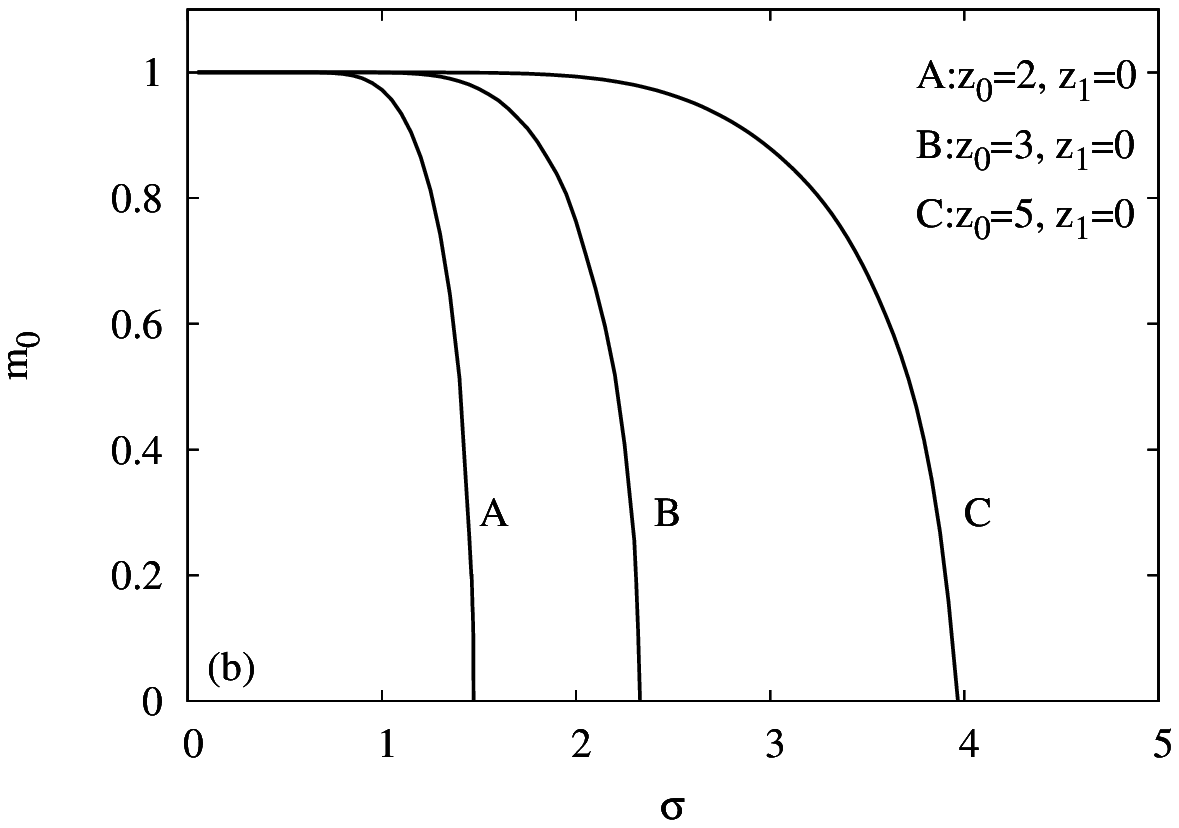, width=6.5cm}
\end{center}
\caption{(a) Phase diagrams of the isotropic quantum Heisenberg model with Gaussian magnetic field distribution in $(k_BT_c/J,\sigma)$ plane for selected lattices. (b) Variation of the ground state magnetization of the system with ($\sigma $) for selected lattices. In (b), the temperature has been fixed as $k_BT_c/J=0.001$.} \label{sek1}\end{figure}

In order to see the effect of the random field distribution width ($\sigma$) on the critical temperature of the model, we depict the variation of the critical temperature  with $\sigma$ in ($k_BT_c/J,\sigma$) plane in Fig. \ref{sek1} for some selected lattices, namely honeycomb ($z_0=2,z_1=0$), square ($z_0=3,z_1=0$) and simple cubic ($z_0=5,z_1=0$) lattices. As we can see from Fig. \re{sek1} that  increasing $\sigma$ values decrease the critical temperature of the whole lattices continuously. This is due to increasing randomness in the system: Increasing the randomness via raising the width of the magnetic field distribution drags the system to the disordered phase while the spin-spin interaction tends to keep the system in an ordered phase. After a critical value of the width, randomness prevails and for the values that provide $\sigma>\sigma_c$, the spin-spin interaction can not maintain an ordered phase even at the zero temperature. At a certain value of $\sigma$, the critical temperature of the simple cubic lattice is greater than that of the square lattice, due to the excess number of nearest neighbors of square lattice. The same relation is also valid between square and honeycomb lattices. Also due to the same reason, simple cubic lattice has a wider ferromagnetic region in a $(k_BT_c/J,\sigma)$ plane than the square lattice and the square lattice has a wider one in comparison with the honeycomb lattice.  Besides, as we can see from Fig. \re{sek1} that the whole phase transitions between ordered and disordered phases are of the second order and the system does not exhibit a reentrant behavior.

In Fig. \re{sek1} (b), we depict the low temperature magnetization versus $\sigma$ curves for the lattices mentioned above. Here, we fixed the temperature as $k_BT/J=0.001$ and this may be considered as the ground state due to the fact that thermal energy supplied by the temperature on the system may be neglected in comparison with the spin-spin interaction energy. The behavior at this temperature is found to be different from the case in the vicinity of the critical temperatures, since increasing $\sigma$ first can not change the ground state magnetization value while it changes the critical temperature continuously. After than, the ground state magnetization starts to reduce, and vanishes with increasing $\sigma$. Due to the significant strength of the spin-spin interaction originating from the large number of nearest neighbors concerning the simple cubic lattice, the ground state magnetization of the simple cubic lattice can become saturated at $1.0$ even for large $\sigma$ values, in contrast to the square and honeycomb lattices. The $\sigma_c$ values at which the critical temperatures reduce to zero corresponding to different lattices can be seen in Table (1). As seen in the Table (1), the EFT-2 formulation is able to distinguish between the lattices with the same coordination number, but different geometry, i.e. the critical width values of the Kagome and square lattices, or triangular and simple cubic lattices are different. For instance, $\sigma_c=2.262$ for Kagome lattice is slightly lower than the same value for the square lattice $\sigma_c=2.320$. Although these two lattices have same number of nearest neighbor, this difference comes from the difference between the Kagome ($z_0=2,z_1=1$) lattice and the square lattice  ($z_0=3,z_1=0$) .

The critical value of $\sigma$ for the 3D lattices can be compared with the same values for the Ising model which are $\sigma_c=3.850$ for the simple cubic, $\sigma_c=5.450$ for the body centered cubic and  $\sigma_c=8.601$ for the face centered cubic lattices\cite{ref21}. For the isotropic Heisenberg model these critical values are slightly greater than the Ising counterparts.

\begin{table}[h]\label{table1}
\begin{center}
\begin{threeparttable}
\caption{The critical value of distribution width ($\sigma_c$) of the isotropic quantum Heisenberg model with zero centered Gaussian magnetic field distribution.}
\renewcommand{\arraystretch}{1.3}
\begin{tabular}{lllllllll}
\thickhline
Lattice & $z_0$& $z_1$ &$\sigma_c (H)$ \\ 
\hline
Honeycomb & 2& 0 & 1.466\\          
Kagome & 2& 1 & 2.262\\             
Square& 3& 0 & 2.320 \\              
Triangular& 3& 2 & 3.875&\\          
Simple cubic& 5& 0 & 3.985&\\        
Body centered cubic& 7& 0 & 5.563&\\   
Face centered cubic& 7& 4 & 8.844& \\   
\thickhline \\
\end{tabular}
\end{threeparttable}
\end{center}
\end{table}

Now let us investigate the evolution of the variation of the magnetization curves with temperature with increasing randomness in the system. In order to achieve this, we depict the variation of the magnetization with temperature curves for some selected values of $\sigma$ in Fig. \re{sek2} for honeycomb and square lattices. As seen in Fig.\re{sek1}, since the difference between the 2D and 3D lattices regarding to the behavior of the critical temperature and ground state magnetization is only quantitative, we do not expect a qualitative difference between the magnetization profiles of 2D and 3D lattices with zero centered Gaussian random field distribution. As seen in Fig. \re{sek2}, magnetization versus temperature curves related to the honeycomb and square lattices are qualitatively similar. As $\sigma$ increases then both critical temperature and the ground state magnetization become reduced. However, $\sigma=1.0$ curve starts with the same value of the $\sigma<1.0$ curves for the square lattice while this is not in the case for the honeycomb lattice. This situation supports the behavior of the ground state magnetizations given in Fig. \re{sek1} (b). Moreover, there is not any observed first order transition in the curves with increasing $\sigma$. This also indicates that increasing $\sigma$ can not induce a first order transition in the isotropic Heisenberg model.

\begin{figure}[h]\begin{center}
\epsfig{file=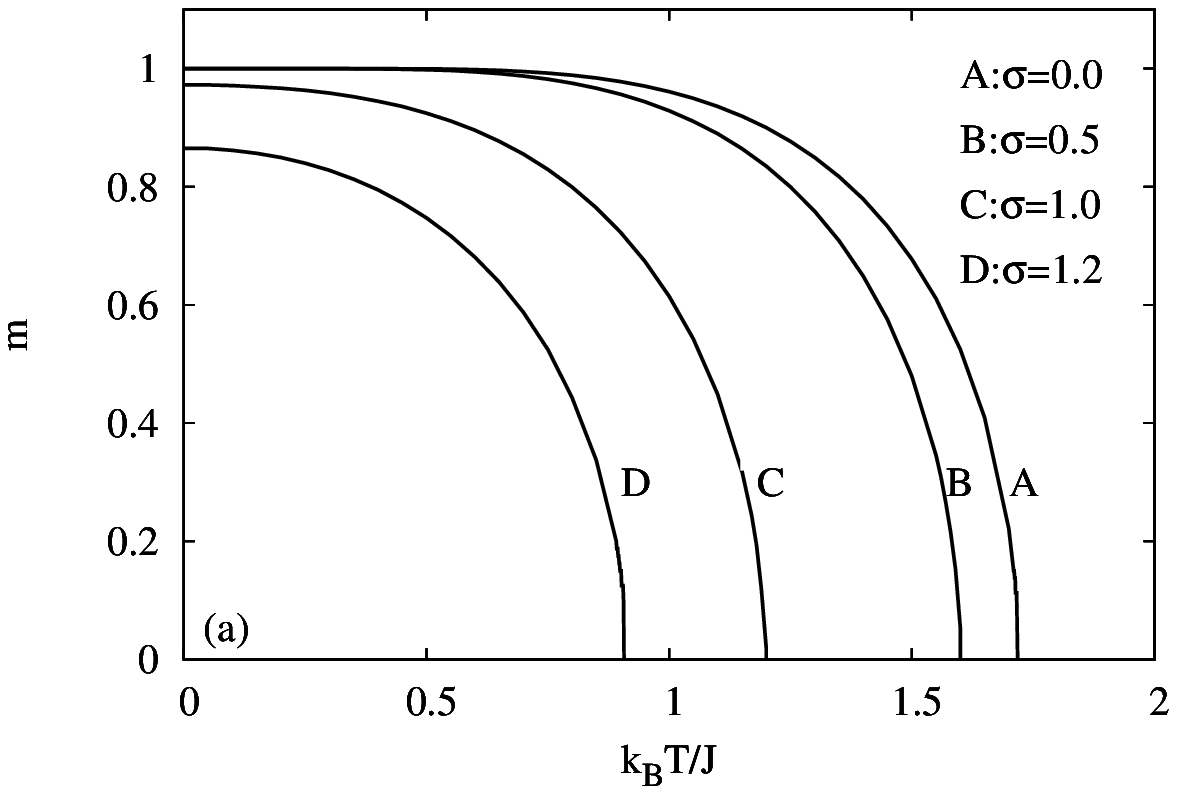, width=6.5cm}
\epsfig{file=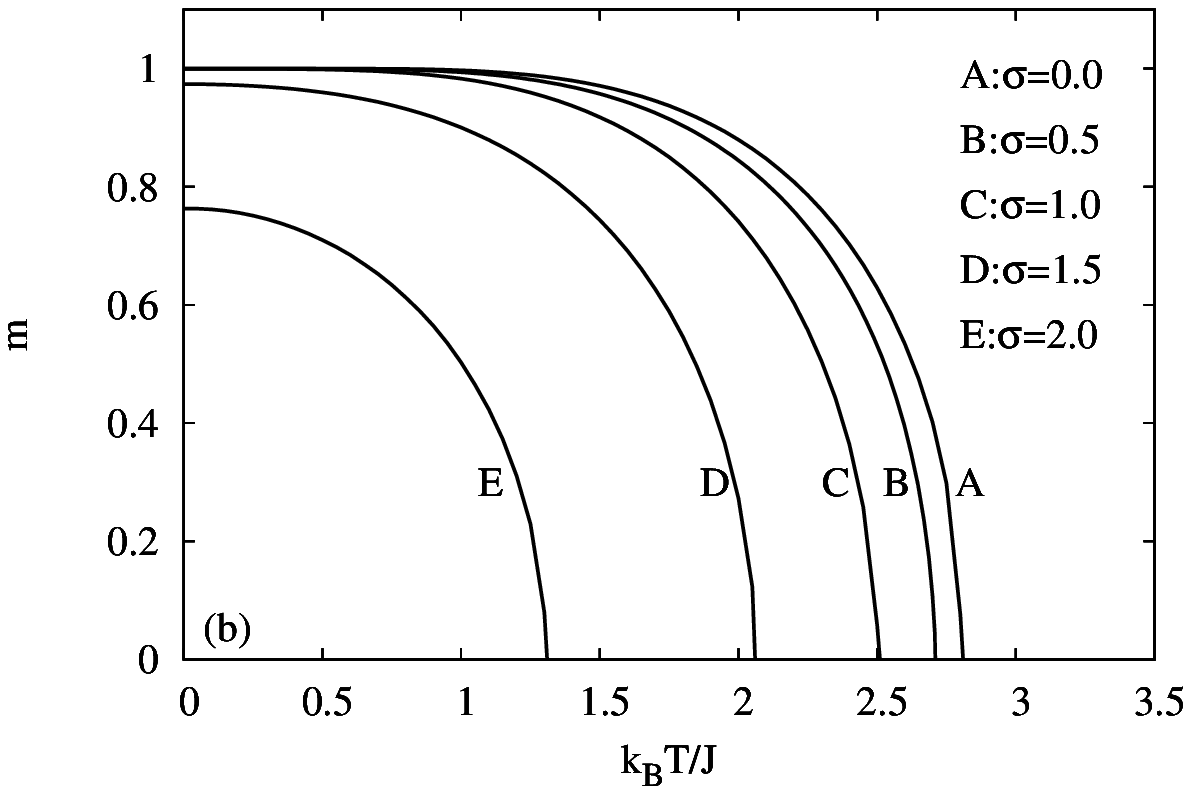, width=6.5cm}
\end{center}
\caption{(a) Phase diagrams of the isotropic quantum Heisenberg model in $(k_BT_c/J,\sigma)$ plane with Gaussian magnetic field distribution for selected lattices. (b) Variation of the ground state magnetization of the system with  ($\sigma $) for selected lattices. In (b), the temperature has been fixed as $k_BT_c/J=0.001$.} \label{sek2}\end{figure}

\section{Conclusion}\label{conclusion}

In conclusion, the effect of the zero centered Gaussian random magnetic field distribution on the phase diagrams of the isotropic quantum Heisenberg model has been investigated in detail. In this regard, the effects of the random magnetic fields have been discussed for 2D and 3D lattices. The phase diagrams, which are the variation of the critical temperature with the width of the Gaussian distribution have been depicted for the honeycomb, square and simple cubic lattices. It has been found that there is not any reentrant behavior or first order transition in the system. The critical Gaussian distribution width at which the critical temperature of the system vanishes has been obtained for several 2D and 3D lattices within the isotropic model.

Besides, the effects of the increasing randomness of the magnetic field distribution on the behavior of the magnetization versus temperature curves have been investigated. We have not observed any qualitative difference between the results for 2D and 3D lattices. We have found that as $\sigma$ increases then the ferromagnetic region in $(m-k_BT_c/J)$ plane becomes narrower and finally ferromagnetic region disappears right after the critical value of the Gaussian distribution width.




\appendix

\section{Symmetry properties of the coefficients}\label{sec_app_a}

In order to see the symmetry properties coefficients given in Eq. \re{denk12}, let us start with writing coefficients given Eq. \re{denk10} as
\eq{denk_app_1}{
C^\prime_{pqr}=\komb{z_0}{p}\komb{z_0}{q}\komb{z_1}{r}\Theta_{pqr}
} where
\eq{denk_app_2}{
\Theta_{pqr}=\integ{}{}{}dH_1dH_2P(H_1)P(H_2)\Theta^\prime_{pqr}\paran{H_1,H_2}
} and

\eq{denk_app_3}{
\Theta^\prime_{pqr}\paran{H_1,H_2}=A_x^{z_0-p}A_y^{z_0-q}A_{xy}^{z_1-r}B_x^{p}B_y^{q}B_{xy}^{r}f\paran{x,y,H_1,H_2}|_{x=0,y=0}.
} Here the function defined by Eq. \re{denk7} and the distribution function for the  magnetic field on the sites labeled by $1$ and $2$ is given by Eq. \re{denk2}. From the definitions given in Eqs. \re{denk4}, \re{denk5} and with using Binomial expansion, \re{denk_app_3} can be written in the form
\eq{denk_app_4}{
\Theta^\prime_{pqr}\paran{H_1,H_2}=\summ{t_1=0}{z_0-p}{}\summ{v_1=0}{p}{}\summ{t_2=0}{z_0-q}{}
\summ{v_2=0}{q}{}\summ{t_3=0}{z_1-r}{}\summ{v_3=0}{r}{}K_{\mathbf{t,v}}f\paran{a_1,a_2,H_1,H_2}
} where $\mathbf{t,v}$ stands for the $(t_1,t_2,t_3,v_1,v_2,v_3)$ and
\eq{denk_app_5}{
\begin{array}{lcl}
a_1&=&\paran{z_0-2t_1-2v_1+z_1-2t_3-2v_3}J_z\\
a_2&=&\paran{z_0-2t_2-2v_2+z_1-2t_3-2v_3}J_z\\
K_{\mathbf{t,v}}&=&\komb{z_0-p}{t_1}\komb{p}{v_1}\komb{z_0-q}{t_2}
\komb{q}{v_2}\komb{z_1-r}{t_3}\komb{r}{v_3}
\\&\times&\paran{-1}^{v_1+v_2+v_3}2^{-\paran{2z_0+z_1}}.\\
\end{array}
}
We can see from Eq. \re{denk_app_4} that, expanded form of the expression have both of the terms which has $f\paran{a_1,a_2,H_1,H_2}$ and $f\paran{-a_1,-a_2,H_1,H_2}$ for all possible values of $a_1$ and $a_2$. In other words, every term which include $f\paran{a_1,a_2,H_1,H_2}$ has a corresponding term $f\paran{-a_1,-a_2,H_1,H_2}$. Let us focus on these two term. We can see from the definitions given in Eq. \re{denk_app_5} that
the transformation
\eq{denk_app_6}{
\begin{array}{lcl}
t_1&\rightarrow&z_0-p-t_1\\
v_1&\rightarrow&p-v_1\\
t_2&\rightarrow&z_0-q-t_2\\
v_2&\rightarrow&q-v_2\\
t_3&\rightarrow&z_1-r-t_3\\
v_3&\rightarrow&r-v_3\\
\end{array}
} transforms the terms in Eq. \re{denk_app_5} as
\eq{denk_app_7}{
\begin{array}{lcl}
a_1&\rightarrow&-a_1\\
a_2&\rightarrow&-a_2\\
K_{\mathbf{t,v}}&\rightarrow&\paran{-1}^{p+q+r-2\paran{v_1+v_2+v_3}}K_{\mathbf{t,v}}\\
\end{array}
} i.e. the terms which have $f\paran{a_1,a_2,H_1,H_2}$ and $f\paran{-a_1,-a_2,H_1,H_2}$ in the expanded form of the Eq. \re{denk_app_4} have same coefficients if $p+q+r$  is even and same but opposite signed coefficients if $p+q+r$  is odd. Thus we can write \re{denk_app_4} as
\eq{denk_app_8}{\Theta^\prime_{pqr}\paran{H_1,H_2}=\left\{
\begin{array}{lcl}
\summ{\mathbf{t,v}}{}{}\frac{K_{\mathbf{t,v}}}{2}\left[f\paran{a_1,a_2,H_1,H_2}+f\paran{-a_1,-a_2,H_1,H_2}\right] &,& \text{p+q+r is even}\\
\summ{\mathbf{t,v}}{}{}\frac{K_{\mathbf{t,v}}}{2}\left[f\paran{a_1,a_2,H_1,H_2}-f\paran{-a_1,-a_2,H_1,H_2}\right] &,& \text{p+q+r is odd}\\
\end{array}\right.
.} Here, the limits of the sums are identical to those in Eq. \re{denk_app_4} and instead of using six sum as in Eq. \re{denk_app_4},
only one sum is used in short notation.

From Eqs. \re{denk7} and \re{denk8} we can see that the function satisfies
\eq{denk_app_9}{
f\paran{a_1,a_2,H_1,H_2}=-f\paran{-a_1,-a_2,-H_1,-H_2}
} then we can write \re{denk_app_8} as
\eq{denk_app_10}{\Theta^\prime_{pqr}\paran{H_1,H_2}=\left\{
\begin{array}{lcl}
\summ{\mathbf{t,v}}{}{}\frac{K_{\mathbf{t,v}}}{2}\left[f\paran{a_1,a_2,H_1,H_2}-f\paran{a_1,a_2,-H_1,-H_2}\right]&,& \text{p+q+r is even}\\
\summ{\mathbf{t,v}}{}{}\frac{K_{\mathbf{t,v}}}{2}\left[f\paran{a_1,a_2,H_1,H_2}+f\paran{a_1,a_2,-H_1,-H_2}\right]&,& \text{p+q+r is odd}\\
\end{array}\right. .
} Thus, we can conclude from Eq. \re{denk_app_10} that,

\eq{denk_app_11}{\Theta^\prime_{pqr}\paran{H_1,H_2}=\left\{
\begin{array}{lcl}
-\Theta^\prime_{pqr}\paran{-H_1,-H_2}&,& \text{p+q+r is even}\\
\Theta^\prime_{pqr}\paran{-H_1,-H_2}&,& \text{p+q+r is odd}\\
\end{array}\right. .
}


Now, if we look at the integrant of Eq. \re{denk_app_2}, with the help of the  Eqs. \re{denk2} and \re{denk_app_11},
we can see that it is symmetric about the origin for the odd valued $p+q+r$  and antisymmetric about the origin for the even valued $p+q+r$
in the $(H_1,H_2)$ plane. Thus, with using this result in Eqs. \re{denk_app_2} and \re{denk_app_1} then in Eq. \re{denk12}   we arrive the property
\eq{denk_app_12}{C_k\left\{
\begin{array}{lcl}
=0&,& \text{p+q+r is even}\\
\ne  0 &,& \text{p+q+r is odd}\\
\end{array}\right.
}
and this completes of our derivation.



\bibliographystyle{model1-num-names}
\bibliography{<your-bib-database>}

\newpage

 \begin{thebibliography}{00}




\bibitem{ref1} A. I. Larkin, Sov. Phys. JETP \textbf{31}, 784 (1970).

\bibitem{ref2} Y. Imry and S. K. Ma, Phys. Rev. Lett. \textbf{35}, 1399 (1975).


\bibitem{ref3}  S. Fishman and A. Aharony, J. Phys. C \textbf{12}, L729 (1979).
\bibitem{ref4}  J. L. Cardy, Phys. Rev. B \textbf{29}, 505 (1984).

\bibitem{ref5}   Daniel S. Fisher, Geoffrey M. Grinsrein, Anil Khurana, Physics Today \textbf{56},
December (1988).


\bibitem{ref6}   M.E. Fisher, J. Chem. Soc. Faraday Trans. \textbf{2}, 1569 (1986).
\bibitem{ref7}   G. Forgacs, R. Lipowsky, T.M. Nieuwenhuizen C. Domb, J. Lebowitz
(Eds.), Phase Transitions and Critical Phenomena, vol. 14 Academic Press,
London (1991), p. 136


\bibitem{ref8}   R. Blossey, T. Kinoshita, J. Dupont-Roc, Physica A \textbf{248}, 247 (1998).

\bibitem{ref9}   Y. El Amraoui, A. Khmou
J. Magn. Magn. Mater. \textbf{218}, 182 (2000).

\bibitem{ref10}  Y. El Amraoui, A. Hamid, S. Sayouri
J. Magn. Magn. Mater. \textbf{219}, 89 (2000).




\bibitem{ref11} A. Aharony, Phys. Rev. B \textbf{18}, 3318 (1978).
\bibitem{ref12} D. C. Mattis, Phys. Rev. Lett. \textbf{55}, 3009 (1985).
\bibitem{ref13} E. F. Sarmento and T. Kaneyoshi, Phys. Rev. B \textbf{39}, 9555 (1989).
\bibitem{ref14} N. G. Fytas, A. Malakis, and K. Eftaxias, J. Stat. Mech. Theory
Exp. (2008), 03015.
\bibitem{ref15} I. A. Hadjiagapiou, Physica A \textbf{389}, 3945 (2010).




\bibitem{ref16} T. Schneider and E. Pytte, Phys. Rev. B \textbf{15}, 1519 (1977).

\bibitem{ref17} T. Kaneyoshi, Physica A 139, 455 (1985).

\bibitem{ref18} Y. Q. Liang, G. Z. Wei, Q. Zhang, Z. H. Xin, and G. L. Song,
J. Magn. Magn. Mater. 284, 47 (2004).

\bibitem{ref19} N. Crokidakis and F. D. Nobre, J. Phys. Condens. Matter 20,
145211 (2008).

\bibitem{ref20} O. R. Salmon, N. Crokidakis, and F. D. Nobre, J. Phys. Condens.
Matter 21, 056005 (2009).

\bibitem{ref21} \"{U}. Ak{\i}nc{\i}, Y. Y\"{u}ksel, H. Polat,
Phys. Rev. E \textbf{83}, 061103 (2011).






\bibitem{ref22} Douglas F. de Albuquerque, A.S. de Arruda
Physica A \textbf{316}, 13 (2002).

\bibitem{ref23} A. Oubelkacem, K. Htoutou, A. Ainane, M. Saber
Chin. J. Phys. \textbf{42}, 717 (2004).

\bibitem{ref24} Douglas F. de Albuquerque, Sandro L. Alves, A.S. de Arruda
Phys. Lett. A \textbf{346}, 128 (2005).

\bibitem{ref25} J. Ricardo de Sousa, Douglas F. de Albuquerque, Alberto S. de Arruda,
Physica A \textbf{391}, 3361 (2012).




\bibitem{ref26} R. Honmura, T. Kaneyoshi
J. Phys. C \textbf{12}, 3979 (1979).


\bibitem{ref27} H.B. Callen
Phys. Lett. \textbf{4}, 161 (1963).
\bibitem{ref28} H. Suzuki
Phys. Lett \textbf{19}, 267 (1965).




\bibitem{ref29} T. Kaneyoshi, Acta Phys. Pol. A \textbf{83}, 703 (1993).


\bibitem{ref30} A. Bob\'{a}k, M. Ja\v{s}\v{c}ur
Phys. Status Solidi B \textbf{135}, K9 (1986).



\bibitem{ref31} T. Idogaki, N. Ury\^{u}
Physica A \textbf{181}, 173 (1992).


\bibitem{ref32} J. Mielnicki,  G. Wiatrowski, T. Balcerzak, J. Magn. Magn. Mater  \textbf{71}, 186 (1988).


\bibitem{ref33} Ijan\'{\i}lio G. Ara\'{u}jo, J. Cabral Neto, J. Ricardo de Sousa
Physica A \textbf{260}, 150 (1998).



\bibitem{ref34} J. Ricardo de Sousa, Douglas F. de Albuquerque
Physica A \textbf{236}, 419 (1997).

\bibitem{ref35} J. Ricardo de Sousa, Physica A \textbf{256}, 383 (1998).

\bibitem{ref36} Y. Miyoshi, A. Tamaka, J.W. Tucker, T. Idogaki
J. Magn. Magn. Mater. \textbf{205}, 110 (1999).


\bibitem{ref37} T. Idogaki, A. Tanaka, J.W. Tucker
J. Magn. Magn. Mater. \textbf{177$-$181}, 157 (1998).


 \end{thebibliography}

\end{document}